\documentclass[amstex,useAMS,epsf,amssymb,usenatbib, twocolumn]{mn2e}
\usepackage{graphicx}
\usepackage{epsfig}
\usepackage{amssymb,amsmath,color}
\usepackage{hyperref}
\usepackage{soul}
\sethlcolor{green}
\usepackage[normalem]{ulem}
\usepackage{url}
\usepackage{breakurl}
\usepackage{tabularx}
\usepackage{natbib}

\newcommand{\fermilat}{{\it Fermi}-LAT}
\newcommand{\gray}{$\gamma$-ray}
\newcommand{\grays}{$\gamma$ rays}

\title[Testing spatial uniformity of the CR spectrum]{Testing spatial
uniformity of the CR spectrum in the local ISM with \gray{} observations}

\author[D. A. Prokhorov \& S. Colafrancesco]{D. A. Prokhorov$^{1}$\thanks{E-mail:dmitry.prokhorov@wits.ac.za} and S. Colafrancesco$^1$\\
~\\
$^{1}$ School of Physics, University of the Witwatersrand, Private
Bag 3, 2050 Johannesburg, South Africa}

\date{\today}

\pagerange{\pageref{firstpage}--\pageref{lastpage}} \pubyear{2018}

\begin{document}

\maketitle

\begin{abstract}
Gamma-ray observations of nearby radio-line-emitting gas structures
in the interstellar medium allow us to probe the spectrum of cosmic
rays (CRs). In this paper, we analysed \textit{Fermi} Large Area Telescope (LAT)
\gray{} observations of three such structures located near each other
to check if their CR spectra are compatible with that of the CR background
or might provide evidence for a population of ``fresh'' CRs.
We found that the shape of the $\gamma$-ray spectrum in the
Aquarius HI shell is consistent with the previously published
stacked $\gamma$-ray spectrum of the Gould Belt molecular clouds. We
also found that assumptions on the diffuse Galactic \gray{}
background affect the spectral shapes of CRs derived in
the R Coronae Australis and $\rho$ Ophiuchi molecular clouds
in which spectral deviations had previously been suggested.
These two facts provide evidence to
support the hypothesis of uniformity of the shapes of cosmic ray
spectra in the local Galaxy environment.
\end{abstract}


\section{Introduction}

Every supernova (SN) explosion releases $\sim 10^{51}$ ergs of
kinetic energy in the interstellar medium (ISM). Stellar material is
ejected at supersonic speeds, generating a shock wave into the ISM.
The ejecta sweep up the surrounding gas and pile up a dense shell.
The SN remnant (SNR) can expand over tens of parsec before the
ejecta speed becomes subsonic. As the ejecta expands, the shocked
gas cools and eventually recombines forming a neutral hydrogen (HI)
shell \citep[for a review, see][]{Reynolds17}. Multiple nearby SN
explosions result in an ensemble of shock waves which inflate a
supperbubble surrounded by a massive HI supershell
\citep[e.g.,][]{McCray87}. SNRs are recognised as the sources of
Galactic cosmic rays (CRs) with energies below $\sim100$ TeV
\citep[for a review, see][]{Berezinsky90}. It is commonly accepted
that CRs gain energy through diffusive Fermi acceleration at fronts
of strong SNR shocks \citep[e.g.,][]{Krymsky77}. The diffusive shock
acceleration produces hard spectra of CRs, $N(E)\propto E^{-2.0}$.
After being accelerated, CRs escape from SNR shocks and travel
through the ISM. CRs are confined in the Galaxy for $\sim10^7$ yr at
GeV energies and during this time the CRs accelerated by individual
sources mix and contribute to the CR background. The diffusive
propagation of CRs in the Galaxy makes the CR spectrum softer due to
energy dependent escape rate. The resulting CR spectrum directly
measured near the Earth is $N(E)\propto E^{-2.8}$ at energies from
20 GeV to 200 GeV \citep[][]{Adriani11, Aguilar15}.

CRs undergo hadronic interactions with the interstellar gas and
produce neutral pions, which decay into \grays{}. The ISM is mainly
composed of hydrogen in atomic or molecular form \citep[for a
review, see][]{Ferriere01}. The tracers of these forms of hydrogen,
the HI 21-cm line \citep{Dickey90, Kalberla05} and the CO 2.6-mm
line, respectively, revealed various structures in the ISM,
including HI shells and supershells \citep[e.g.,][]{Heiles79, Hu81,
Ehlerova13} and molecular clouds \citep[MCs; for a review,
see][]{Heyer15}. SNs are expected to occur near the molecular
material in which their massive progenitor stars were born. Exploding in
the vicinity of MCs, SNs make HI shells and MCs nearby objects,
while CRs escaping from SNRs can interact with the MC gas producing \grays{}
\citep[e.g.,][]{Montmerle79, Uchiyama12, Aharonian08}.

If one SN occurs within 100 pc distance every Myr \citep[that is the
average SN rate in the Galactic disk, e.g.,][]{Diehl06}, the
injection of CRs into the ISM can leave an imprint on the CR spectra
local to the injection site \citep[e.g.,][]{Kachelriess15}. It was
proposed by \citet[][]{Neronov17} that discreteness of CR injection
by SN events in space and time can result in the variations of the
slope of the CR spectrum in the energy range of 10 GeV to 1 TeV
across the local ISM. In this scenario, the CR spectrum has a
low-energy break and the slope of the CR spectrum below the break
energy is determined by the balance of the CR steady-state injection
and energy-dependent propagation, whereas the variation of the
spectrum above the break energy is attributed to the gradual
transition from the steady-state continuous injection to the regime
of discrete source injection. To test this scenario, they analysed
$\gamma$-ray observations accumulated by \textit{Fermi} Large Area
Telescope (LAT) of high-latitude MCs belonging to the Gould Belt and
reported that the slope of CR spectrum above 18$^{+7}_{-4}$ GeV is
variable across the 1 kpc scale region, but stated that the
statistics of \textit{Fermi}-LAT data is only marginally sufficient
to establish the two regimes of CR injection. Other studies of the
CR hadron spectrum within 1 kpc used \textit{Fermi}-LAT observations
of a mid-latitude region in the third Galactic quadrant
\citep[][]{local09} and of the nearby MBM 53, 54, and 55 MCs and the
far-infrared loop-like structure in Pegasus located within 100-150 pc
of the Sun \citep[][]{Mizuno16}.
The latter CR spectrum is seemingly harder than that from other
\fermilat{} studies \citep{local09, jm15}.

The ejecta from a local SN explosion occurred at the distance
$\sim100$ pc has apparently reached the Earth some 2 Myr ago
\citep[][]{Knie04, Wallner16}. The same SN might also lead to
variation in the CR spectra of the Gould Belt MCs located within
$\sim$200 kpc. The three nearest MCs amongst the high-latitude
Gould Belt MCs include $\rho$ Ophiuchi (Oph), R Coronae Australis (R
CrA)\footnote{the R CrA MC does not lie inside the Gould Belt}, and
Taurus located at distances of 165 pc, 150 pc, and 140 pc,
respectively. The $\gamma$-ray spectrum of the $\rho$ Oph MC
is the hardest amongst the high latitude Gould Belt MCs, while the $\gamma$-ray
spectrum of the R CrA MC differs from the spectra of the other MCs
from the sample of \citet[][]{Neronov17}. We compared mutual
distances between the MCs from that sample and found that R CrA and
$\rho$ Oph MCs are separated by only 100 pc. This mutual distance
is the smallest in the sample with the exception of the two MCs, Orion A and Orion B,
in the Orion complex located 500 pc away. Given the peculiar CR
spectral shapes in the $\rho$ Oph and R CrA MCs and their
short mutual distance as well as the possible CR spectral hardness of the
nearby MBM 53-55 MC and the Perseus loop, it makes sense to performed a detailed analysis of
the spectra of the $\rho$ Oph and R CrA MCs and
of the CR spectrum of another structure located in their
vicinity in order to search for a signature of discrete source CR injection.
The Aquarius (Aqr) shell identified using the HI 21-cm
line observation by \citet[][]{Hu81} and using far-infrared emission
by \citet[][]{Konyves07} is located at a distance of 170 pc
\cite[][]{Hu81}.
If we adopt this distance then the Aqr shell is at distances
of 110 pc and 190 pc from the R CrA and $\rho$ Oph MCs,
respectively, and can provide us with an alternative probe of the CR
spectrum in their neighbourhood.

In this paper, we analyse \textit{Fermi}-LAT observations of the Aqr
HI shell and the R CrA and $\rho$ Oph MCs to test uniformity of the
CR spectrum in the local Galaxy.
In Section~2, we discuss the \textit{Fermi}-LAT observations that we
used and their analysis. We describe our models for the $\gamma$-ray
emission of these nearby ISM structures in Section~3 and we discuss our
results in Section~4. We discuss prospects of indirect measurements
of the CR spectra in ISM gas accumulations located at short mutual distances
in Section~4 and finally present our conclusions in Section~5.

\section{\textit{Fermi}-LAT observations and analysis}

Gamma-ray observations of radio-line-emitting ISM structures allow
us to indirectly measure CR spectra at different locations in our
Galaxy \citep[e.g.,][]{local09, Casanova10, fermiclouds12, Yang16},
since (i) the Galaxy is transparent for $\gamma$ rays, (ii)
$\gamma$-ray production cross-section is independent of the chemical
or thermodynamic state of the ISM gas, and (iii) $\gamma$-ray
spectra are determined by their parent CR spectra.

The \textit{Fermi}-LAT is a pair-conversion wide field-of-view imaging telescope
onboard the \textit{Fermi} satellite launched in June 2008.
\textit{Fermi}-LAT covers
the $\gamma$-ray energy range from $\sim$20 MeV to several hundreds
of GeV \citep[][]{Atwood09}. To take advantage of its large field-of-view,
the main observing mode of \textit{Fermi} is the sky-survey mode,
which applies that the exposure is almost uniform over the sky after
two \textit{Fermi} orbits (or about 3 hours). It makes possible our
investigation because the $\rho$ Oph and R CrA MCs
and the Aqr shell are projected onto different regions of the sky despite
their close mutual distances.
The \textit{Fermi}-LAT has an angular resolution per single event of 3$^{\circ}$
at 200 MeV, narrowing to 0.8$^{\circ}$ at 1 GeV, and further narrowing to
0.15$^{\circ}$ at 10 GeV \citep[][]{Atwood13}. The LAT instrument detected
more than one billion of $\gamma$-rays, i.e. hundreds of times more
than that detected by the previous-generation EGRET instrument.
Wide energy coverage and large photon statistics provided by \textit{Fermi}-LAT
are crucial for studying the spectra of faint, extended $\gamma$-ray sources,
including those of the Aqr shell and the R CrA MC.

We downloaded the \textit{Fermi}-LAT Pass 8 data from the Fermi
Science Support
Center\footnote{\burl{https://fermi.gsfc.nasa.gov/cgi-bin/ssc/LAT/LATDataQuery.cgi}}.
We selected Pass 8 CLEAN class data (evclass = 256) spanning 9 years
(MET 239557417-521539208) with energies between 200 MeV and 200 GeV.
The low-energy bound is selected to reduce the possible
contamination of the spectral modelling by an electron
bremsstrahlung component that is expected to appear at lower
energies \cite[][]{Neronov12}. The CLEAN class has a lower particle
background rate above 3 GeV than that of the parent SOURCE class and
is better suited for studying the shape of spectra above a few GeV.
We chose two regions of interest (ROI) with a radius of
22$^{\circ}$ that enclose the Aqr shell and the R CrA MC,
respectively, and one ROI with a radius of 15$^{\circ}$ that
encloses the $\rho$ Oph MC. The choice of the radius of the ROI depends on
a projected distance from each of these ISM structures to the Galactic
plane. For the data analysis, we used the \texttt{FERMI
SCIENCE TOOLS} v10r0p5
package\footnote{\burl{https://fermi.gsfc.nasa.gov/ssc/data/analysis/software/}}
and P8R2\_CLEAN\_V6 instrument response functions. To avoid
contamination from the $\gamma$-ray-bright Earth's limb we removed
all events with zenith angle $>90^{\circ}$. We applied the
recommended time selection quality cuts (DATA QUAL==1 \&\& LAT
CONFIG==1), ensuring that the LAT instrument was in normal science
data-taking mode. We binned the data in 30 logarithmically spaced
bins in energy. We used the spatial binning with a pixel size of
$0\fdg15$. To model \gray{} emission in the analysis regions, we
included \gray{} sources from the LAT 4 year point source (3FGL)
catalogue \citep[][]{3FGLcat}. To compute the spectral parameters of
\gray{} sources, we performed a binned likelihood analysis by using
the pyLikelihood
package\footnote{\burl{https://fermi.gsfc.nasa.gov/ssc/data/analysis/scitools/python\_usage\_notes.html}},
which is part of the \texttt{FERMI SCIENCE TOOLS}.

\begin{figure*}
\centering
\includegraphics[angle=0,width=9.5cm]{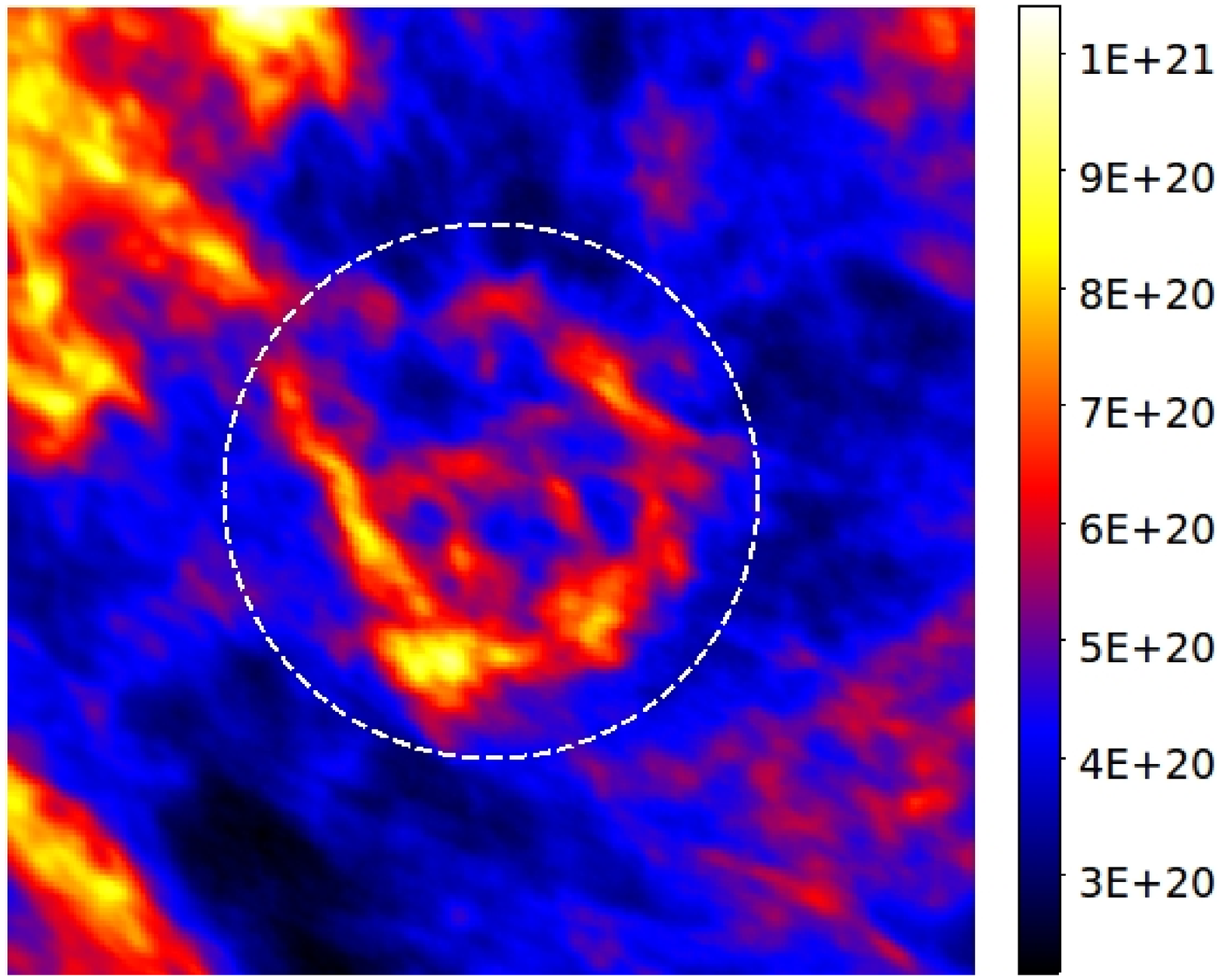}
\includegraphics[angle=0,width=8.0cm]{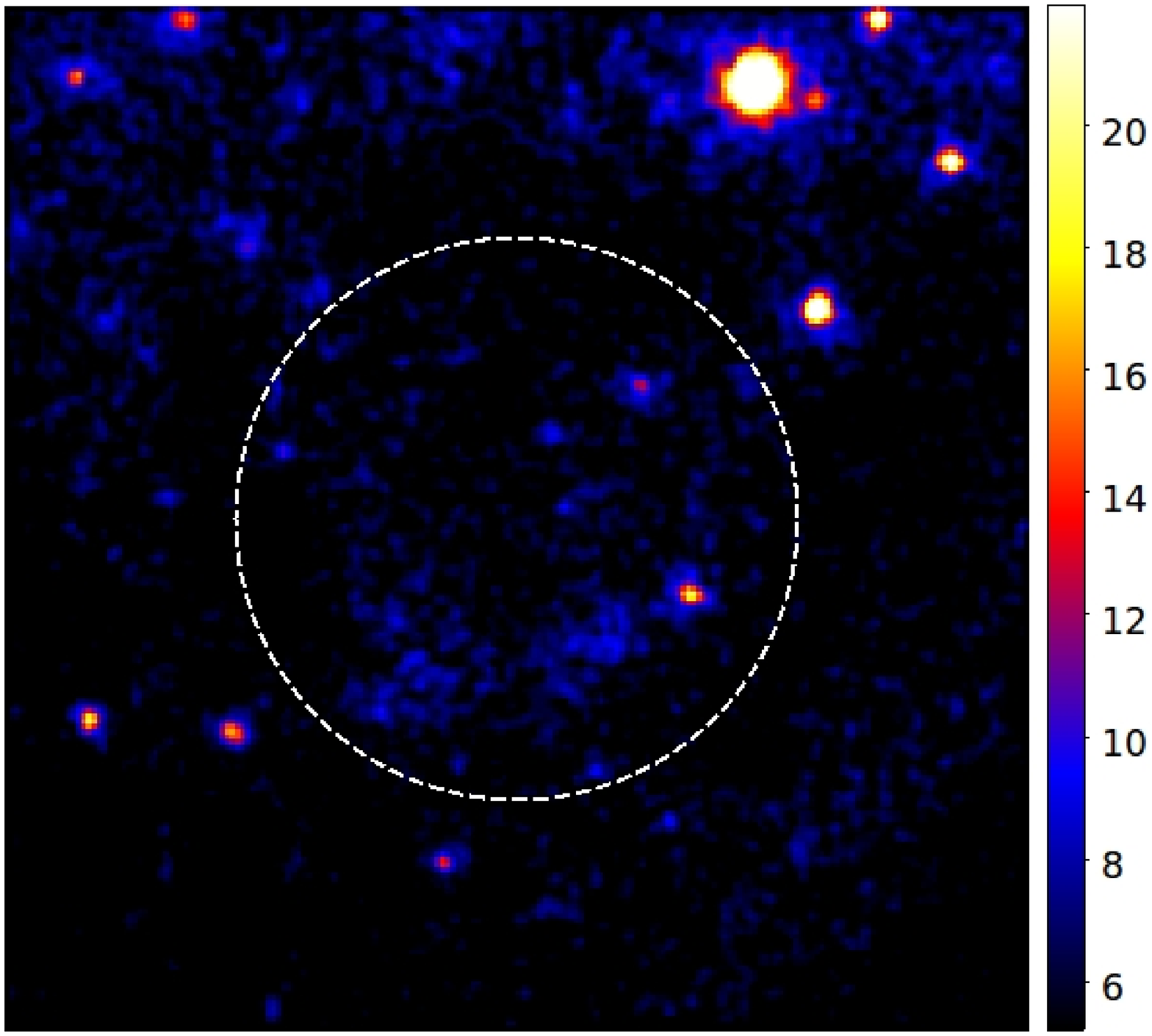}
\caption{The neutral atomic hydrogen column density map of and the
\textit{Fermi}-LAT 0.5-200 GeV count map of the square region of
size $20^{\circ}\times20^{\circ}$ enclosing the Aqr shell
are shown in Galactic coordinates on the left-side and right-side panels,
respectively.}
\label{F1}
\end{figure*}

\section{Model preparation}

Most of the celestial \grays{} detected by \fermilat{} are created
from the interactions of CRs with the ISM. Analyses of faint,
extended \gray{} sources require models of the Galactic diffuse and
isotropic emission. The standard Galactic interstellar emission
model (IEM) is recommended by the \fermilat{} collaboration and is
based on a linear combination of maps for interstellar gas column
density and for the inverse-Compton emission produced in the Galaxy
and also includes large-scale \gray{} emitting structures, such as
Loop I and the Fermi bubbles. To account for dark neutral
gas in our Galaxy, which is not derived from the HI and CO data
\citep[][]{Grenier05}, a template related to the total dust column
density is included in the standard IEM. The IEM we used is
described by the \textit{gll\_iem\_v06.fits} template, released with
Pass 8 data \citep[][]{giem}. The IEM template comprises a set of
all-sky maps representing the expected foreground emission at given
energies. The corresponding isotropic component is described by the
\textit{iso\_P8R2\_CLEAN\_V6\_v06.txt} template.

Two classes of extended \gray{} sources are possible:\\
(a) Extended sources emitting \grays{} via mechanisms
not correlated with the HI density: For modelling of such sources
one needs to include the template corresponding to expected emission
into the source model. This technique was used for detection of the
giant outer lobes of Centaurus A \citep[][]{fermiCenAlobes} and
of extended \gray{} emission from Fornax A \citep[][]{fermiFornaxA}, and
also for searches for new potential extended \gray{} sources,
such as galaxy clusters \citep[e.g.,][]{Han12},
cosmological shock waves \citep[e.g.,][]{Prokhorov14a} or pair haloes
\citep[e.g.,][]{occ14}.\\
(b) Extended sources with \gray{} emission
correlated with the HI density: In this case,
the standard IEM should carefully be refined and such extended
\gray{} sources should be extracted from the IEM template
for modelling\footnote{In preparation of the standard IEM,
any signal from the Magellanic clouds which correlate with the HI density
were extracted}. This class includes Galactic objects, such as
Gould Belt MCs and HI gas shells
\citep[e.g.,][]{Yang14, Mizuno16}, and extragalactic objects,
such as the gaseous disc of the Andromeda galaxy \citep[][]{andromeda10}.
To refine the standard IEM model for objects of this class,
one needs to redo a decomposition of \gray{} emission into the linear
combination of templates for various components of the Galactic
diffuse emission \citep[][]{diffuse2012} or to adopt a background model
from other regions using the standard IEM.
The former method was applied by \citet[][]{Mizuno16} using the
dust properties derived from the \textit{Planck} data to the analysis
of \textit{Fermi}-LAT data. Since the \textit{Planck} data
were not available for the development of the current, stardard IEM,
the former method in this case is advantageous over the latter method.
In this paper, we used the latter method following \citet[][]{Neronov17}
for ease of comparison with their results and elaborated further
on implementation of this method for more detailed background modelling
in its framework.

\subsection{Aquarius HI shell}

The Aqr shell is a Galactic HI structure clearly visible as a large
ring-shaped intensity enhancement in the standard IEM model
template. To illustrate that the HI shell in Aqr is identified with
an extended \gray{} source, we show the neutral atomic hydrogen
column density map in units of cm$^{-2}$ obtained with the survey,
HI4PI \citep[][]{HI4PI16}, and downloaded from the website,
$\burl{https://skyview.gsfc.nasa.gov/}$, on the left-hand side panel
of Fig. \ref{F1} and show the \fermilat{} 0.5-200 GeV\footnote{the
lower bound of 0.5 GeV is selected in order to sharpen the \gray{}
count map in Figure, since the PSF is superior at these energies}
count map smoothed by a Gaussian kernel of $\sigma=0.2^{\circ}$ on
the right-hand side panel. Both the panels map a square of side
20$^{\circ}$ and are centered on ($l$, $b$)=(42$^{\circ}$,
-33$^{\circ}$). The white dashed circles in Fig. \ref{F1} have a
radius of 5.5$^{\circ}$ and encompass the Aqr shell. This shell is
clearly visible on both panels. Using the data of the
Leiden/Argentine/Bonn survey \citep[][]{Kalberla05} downloaded from
the
website\footnote{\burl{https://lambda.gsfc.nasa.gov/product/foreground/fg\_LAB\_HI\_Survey\_get.cfm}},
we checked and found that the contribution of the Aqr shell to the
total HI column density within the central circle of 5.5$^{\circ}$ radius
is dominant ($\gtrsim50\%$) within the velocity range from 0 to +10
km s$^{-1}$ in the local standard of rest (LSR) frame. Looking for
substructures within other velocity ranges, we found that there is
another substructure projected onto the Aqr shell and centered on
($l$, $b$)$\approx$(45$^{\circ}$, -36$^{\circ}$) within the velocity
range from -10 to 0 km s$^{-1}$ in the LSR frame. We downloaded the
CO line emission
map\footnote{\burl{http://pla.esac.esa.int/pla/aio/product-action?MAP.MAP\_ID=COM\_CompMap\_CO-commander\_0256\_R2.00.fits}}
produced from the Planck data and found an CO line emission
counterpart of this substructure. Using the Planck Catalog of
Galactic cold clumps \cite[][]{planckXXVIII}, we found that the
source of PGCC G45.16-36.19 is at this location. The distance to
this cold clump is of 295 pc as taken from \citet[][]{planckXXVIII}.
Taking the contribution from this source into the \gray{} signal
from the Aqr shell, it is possible to set a conservative upper limit
on the distance between the extended \gray{} source towards the Aqr
shell and the R CrA or $\rho$ Oph MCs. These limits are less than
200 pc and 280 pc, respectively.

Before starting an analysis of the $\gamma$-ray signal from this
shell one needs to remove the Aqr shell's component from the
standard IEM template. To refine the template, we replaced in each
energy band the template emission in the circle of $5.5^{\circ}$ radius
encompassing the Aqr shell centered on ($l$, $b$)=(42$^{\circ}$,
-33$^{\circ}$) with a constant value determined as a mean over the
circle of the same size located in the neighborhood foreground
region centered on ($l$, $b$)=(31$^{\circ}$, -33$^{\circ}$). A similar
procedure was used by \citet{Neronov17} in their analysis of the MCs.
We found that this procedure is reasonable at high Galactic
latitudes and if the source is not projected onto the Fermi bubbles.
To avoid the presence of pixels with negative values in
the shell template, the replacement is applied only for the pixels
with initial values higher than the constant value derived from the
neighbourhood foreground region. To produce the template for the Aqr
shell, we subtracted the refined IEM template from the standard IEM
template. The model also includes 3FGL sources which were incorporated
using the make3FGLxml.py script\footnote{the make3FGLxml.py script
automatically adds 10 degrees to the ROI to account
for sources that lie outside the data region}. The three brightest
3FGL $\gamma$-ray point sources which belong to a circular region
of $5.5^{\circ}$ radius surrounding the Aqr shell are 3FGL
J2103.7-1113, 3FGL J2051.3-0828, and 3FGL J2110.3-1013. Their fluxes
above 1 GeV are $1.1\times10^{-9}$, $6.6\times10^{-10}$, and
$3.9\times10^{-10}$ ph cm$^{-2}$ s$^{-1}$ (taken from the 3FGL
catalogue). The 2nd and 3rd of these 3FGL sources are identified with
the pulsar, PSR J2051-082, and the FSRQ blazar, PKS 2107-105,
respectively, while the 1st source is unidentified. We checked the
quality flag for this unidentified \gray{} source in the 3FGL
catalogue and found no caution regarding the reality of this source or
the magnitude of its measured properties. The brightest source in
the entire ROI, 3FGL J2025.6-0736, is located at a distance of
$10^{\circ}$ from the ROI's centre and has a flux of
$7.0\times10^{-9}$ ph cm$^{-2}$ s$^{-1}$ at $E>1$ GeV.
We evaluated the normalisations of the refined IEM component and of
the isotropic component in the ROI using the data. We took the
spectral shapes of the \gray{} point sources from the 3FGL catalogue
and derived the normalisations of the nine brightest sources in the ROI
and of the 3FGL sources overlaid on the Aqr shell template using the
data. To study the spectrum of the Aqr shell, we binned the Aqr
shell's model into bins in energy. The flux normalisation was left
free in each of the energy bins for the Aqr shell component.

\subsection{R CrA molecular cloud}

The R CrA MC is one of the high-latitude MCs.
It is located at significantly lower Galactic latitudes than those of
the Aqr shell and therefore its analysis requires more detailed background
modelling.
This MC is enclosed in the square region of side 6$^{\circ}$ and
with the centre at the Galactic coordinate of (l, b)=($0.56^{\circ}$,
$-19.63^{\circ}$). To model a background at these Galactic
latitudes, one needs to take the diffuse Galactic emission gradient
with latitude into account and to produce a background template varying
with Galactic latitude. The region enclosing the R CrA MC is
projected onto the southern Fermi bubble and near the boundary of
the cocoon, i.e. the region of enhanced \gray{} emission in the
south-east side of the bubble. The \gray{} spectrum of the Fermi
bubbles is hard with index of -2 \citep[][]{Su10} and
is significantly harder than the \gray{} spectrum
of the MC at energies above a few GeV.
Thus, the region selected for background extraction
should have the contribution of \grays{} from the Fermi bubbles
similar to that is expected in the region of the R CrA MC.
The infrared loop centered at (l, b)=($7^{\circ}$, $-20^{\circ}$)
with a radius of $\sim5-6^{\circ}$ \citep[][]{Konyves07} is associated
with the R CrA MC, but its \gray{} emission whether is much fainter than
that of the R CrA MC or is projected onto the bright part of the southern
Fermi bubble's cocoon. Therefore, we derived the \gray{} spectrum for the
R CrA MC, but not for the associated infrared loop.

\begin{figure}
\centering
\includegraphics[angle=0,width=9.0cm]{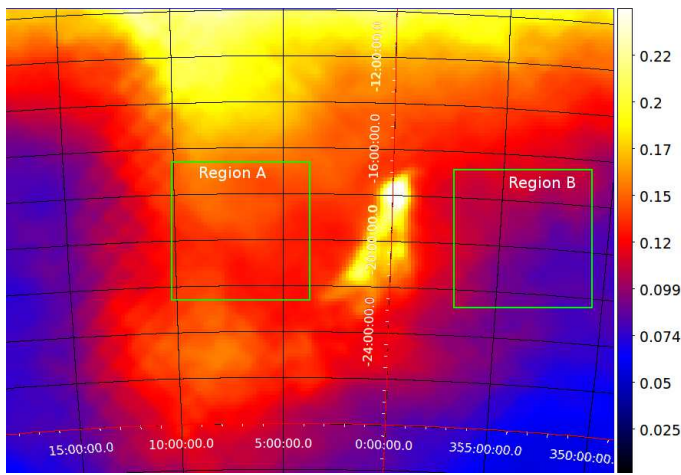}
\caption{The background extraction regions, A and B, overlaid on the
Galactic diffuse \gray{} model count map centered around 9 GeV. The bright \gray{}
emitting structure between these two regions is the R CrA MC.} \label{F2}
\end{figure}

To confidently model the background emission associated with the
southern Fermi bubble, we took two different regions, A and B,
centered at the same Galactic latitude. The region B centered
at (l, b)=($353.93^{\circ}$, $-19.63^{\circ}$) is
shifted compared with the region A centered at (l, b)=($6.94^{\circ}$,
$-19.63^{\circ}$) along the axis of Galactic
longitude. Both the regions have a square box shape with a size of
6$^{\circ}$. The region A is the same as that was used by
\citet[][]{Neronov17} for background extraction and is located in
the cocoon of the Fermi bubble, whereas the region B is located
outside of the cocoon. Therefore, the contributions of \grays{} from
the southern Fermi bubble to these extraction regions are different.
We illustrated it by showing the regions of A and B in Figure \ref{F2}
as well as the Galactic \gray{} model count map centered at energy of 9 GeV.
The \gray{} enhancement caused by the cocoon is clearly visible in
the region A.

To take the diffuse Galactic \gray{} emission gradient into account,
we used the background maps for the regions of A and B adopted from
the standard IEM mapcube template. We scanned these maps along the
Galactic longitude to compute the average value in each latitude
band of 0.125$^{\circ}$ in width. We used these longitude-averaged
values to refine the background IEM model in the square region
enclosing the R CrA MC by replacing in each energy band the IEM
emission with the computed average values. To produce the R CrA MC
template, we subtracted the refined diffuse emission templates from
the standard IEM template. We labelled the model including the
diffuse and R CrA MC templates derived using the background
extraction region, A, as Model A, and the model based on the
background extraction region, B, as Model B. To study the spectrum
of the R CrA MC, we used the R CrA MC template models binned in
energy. To model point \gray{} sources in the analysis region, we
adopted their spectral shapes from the 3FGL catalogue and left the
normalisations of ten bright point \gray{} sources, the refined IEM
component, and the isotropic component to vary freely. Two point
\gray{} sources are overlaid on the R CrA MC template, including
3FGL J1913.5-3631 associated with a \gray{} blazar, PMN J1913-3630,
and 3FGL J1902.3-3702c, which has no clear association. The ``c"
suffix in the latter 3FGL source shows that this 3FGL source is on
top of an interstellar gas clump or that small-scale defect is in
the standard IEM model of diffuse emission, while its significance
in the 3FGL catalogue, $4.7\sigma$, is less than specified minimum for
a free source of 5.0$\sigma$. We also set the normalisations free
for these two sources.

\subsection{$\rho$ Ophiuchi molecular cloud}

The $\rho$ Oph MC lies in the Gould Belt of the closest molecular
complexes. The \gray{} flux from the $\rho$ Oph MC exceeds the
fluxes from the Aqr HI shell and the R CrA MC by factor of $\sim 8$ and
$\sim 5$, respectively. The $\rho$ Oph MC is located at lower Galactic
latitudes of $b\approx16^{\circ}$ and is projected onto the northern
Fermi bubble. Therefore, one needs to take both the diffuse Galactic
\gray{} emission gradient and the hard spectral contribution from
the Fermi bubble into account. \citet[]{Yang14} explains the
observed spectral hardness of the $\rho$ Oph MC by the fact that
point \gray{} sources contribute to the total \gray{} signal and
therefore one needs to consider how to incorporate the point sources
in the model in order to alleviate this problem. Below we list the
improvements over the previous work based on the background
extraction method.

1) We selected the $\rho$ Oph MC region with a size of
$6^{\circ}\times5.5^{\circ}$ centered on ($l$,
$b$)=($353.3^{\circ}$, $16.2^{\circ}$). This selection is based on
the \textit{Planck} maps at frequencies of 28.5 and 44.1 GHz
dominated by dense molecular gas \citep[][]{PlanckXX}. The selected
region has a smaller surface area than that of
$10^{\circ}\times10^{\circ}$ which was previously analysed by means of
the extraction region method. Since the contribution of \grays{}
from the Fermi bubble is expected to scale with surface area, this
contribution is expected to decrease by a factor of 3 in our
analysis.

\begin{figure}
\centering
\includegraphics[angle=0,width=9.0cm]{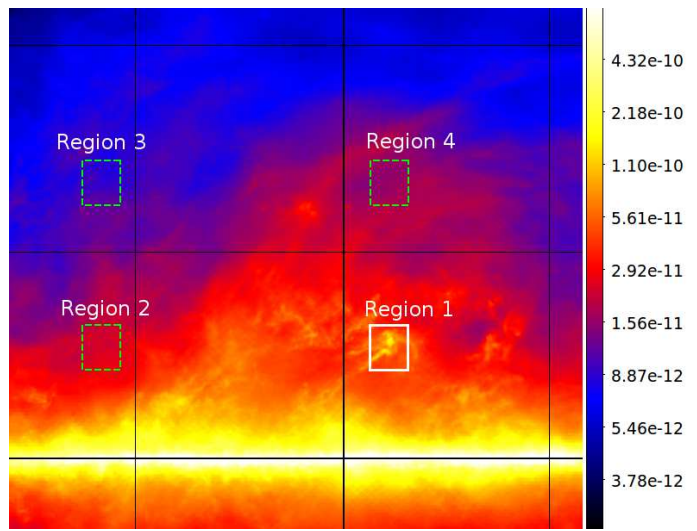}
\caption{The background extraction regions 2, 3, and 4, overlaid on the
Galactic diffuse \gray{} intensity at 9 GeV. The bright \gray{}
emitting structure in the region 1 is the $\rho$ Oph MC.
Grid lines of latitude parallels and longitude meridians
at $30^{\circ}$ intervals are shown.} \label{FigOph1}
\end{figure}

2) We used the first 4 years of \fermilat{} data to perform an
analysis of the $\rho$ Oph MC. Taking into account that its \gray{}
emission is much stronger than those of the other two structures
discussed above, it is reasonable to select the data set described
by the LAT 4 year point source catalogue \citep[][]{3FGLcat} and also
by the diffuse Galactic \gray{} emission template \citep[][]{giem}.
It allows us to adopt the normalisation of the sources from the
catalogue.

3) To model the diffuse Galactic \gray{} gradient, we used the
background extraction region 2 shown as in Fig. \ref{FigOph1}. The
region 2 is shifted by $30^{\circ}$ compared with the region 1,
enclosing the $\rho$ Oph MC, along the axis of Galactic longitude.
We scanned the region 2 along the Galactic longitude to compute the
average value in each latitude band. We used the longitude-averaged
values to refine the background IEM model by replacing in each energy band
the IEM emission in the $6^{\circ}\times5.5^{\circ}$ rectangular
region with the computed average values. We used the refined
background model, in turn, to produce the $\rho$ Oph MC template. We
called this background extraction model Model O1.

4) To account for \grays{} produced in the Fermi bubble which
contaminate the \gray{} signal from the $\rho$ Oph MC, we introduced
two other high-latitude background extraction regions, namely the
regions 3 and 4 (see Fig. \ref{FigOph1}). We subtracted the \gray{}
signal of the region 3 from that of the region 4 to compute the
average contribution of the northern Fermi bubble to \gray{}
emission. We added the northern Fermi bubble's contribution to the
refined background IEM model obtained from the region 2 to further
refine the background model. We also produced the $\rho$ Oph MC
template on the basis of this background model. The contribution
from the northern Fermi bubble to this $\rho$ Oph MC template is
therefore minimised. We called this background extraction model
Model O2.

\section{Results}

In this Section, we present the results of our spectral analyses of
the Aqr shell, the R CrA MC, and the $\rho$ Oph MC based on the
models described above.

To compute the statistical significance of the presence of an
extended \gray{} source at the location of each of these ISM
structures, we used the Test Statistic (TS) defined as
$\mathrm{TS}=-2\ln(L_{\mathrm{max},0}/L_{\mathrm{max},1})$, where
$L_{\mathrm{max},0}$ is the maximum likelihood value for a model
without an additional source (the `null hypothesis') and
$L_{\mathrm{max},1}$ is the maximum likelihood value for a model
with the additional source at a specified location. We chose the
energy bins centered at 240 MeV, 450 MeV, 890 MeV, 1.55 GeV, 2.85
GeV, and  6.5 GeV to yield confident detections in all bins.

We used the \texttt{naima} package \citep[][]{naima15} to search for
suitable estimators characterising the shapes of CR spectra below or
above the energy break using the derived \gray{} fluxes. \texttt{Naima}
is an open source Python package which implements radiative models
for computing the non-thermal emission of relativistic particle distributions,
including hadronic $\gamma$-ray emission.
Given that observed \grays{} were produced in p-p interactions followed by
neutral pion decay \cite[][]{Kafexhiu14}, we fixed the differential energy
spectrum of protons at the energy break of 18 GeV,
the power-law CR spectral index below the break at the value of 2.33, and
allowed the power-law index above the energy break to vary in the range of (2.5, 5.0).
We found that the weighted mean ratio of the fluxes computed with \texttt{naima}
to the stacked MC's modelled fluxes over the first three energy bins,
$R_{1-3}$, only slightly (within 10\%) changes with the high-energy
power-law index value, while the ratio of the flux computed with
\texttt{naima} to the stacked MC's flux in the 6th energy bin (which
is centered at 6.5 GeV) varies with high-energy power-law index by a
factor of 5. Therefore, we introduced these two estimators
allowing us to characterise the low-energy and high-energy parts
of CR spectra, respectively. Note that if two sources have identical
proton energy spectra above the break then the values of both the low- and
high-energy estimators are the same.
The absolute values of estimators are proportional to the ratio
of $\gamma$-ray fluxes of two sources. If two sources have different proton energy
spectra above the energy break, the values of low- and high-energy estimators
are different.

\subsection{Aquarius HI shell}

Using the introduced spectral-spatial model, we performed a binned
likelihood analysis to compute the differential fluxes of the Aqr
shell. We introduced a higher energy bin centered at 16.4 GeV in
addition to the six energy bins. We found that the emission from the
Aqr shell is detected in the first six bins with TS of 158, 245,
244, 158, 68, and 31. The square root of the TS is approximately
equal to the detection significance for a given source. Thus, the
\gray{} emission from the Aqr shell is detected with high confidence
of $>10\sigma$ in each of the first four energy bins and with
5.5$\sigma$ confidence in the 6th energy bin. We found that the
normalisation of the differential flux in each bin is consistent
with that is expected from the model described by the Aqr shell
template if the same scale factor is applied for all these energy
bins. Taking this consistency into account, we fixed the flux scale
factor in each energy bin to the identical value, computed the flux
scale factor value by means of a likelihood method, and found that
the Aqr shell is detected in \grays{} at TS=1187, i.e. with
confidence of 34$\sigma$.

\begin{figure}
\centering
\includegraphics[angle=0,width=9.5cm]{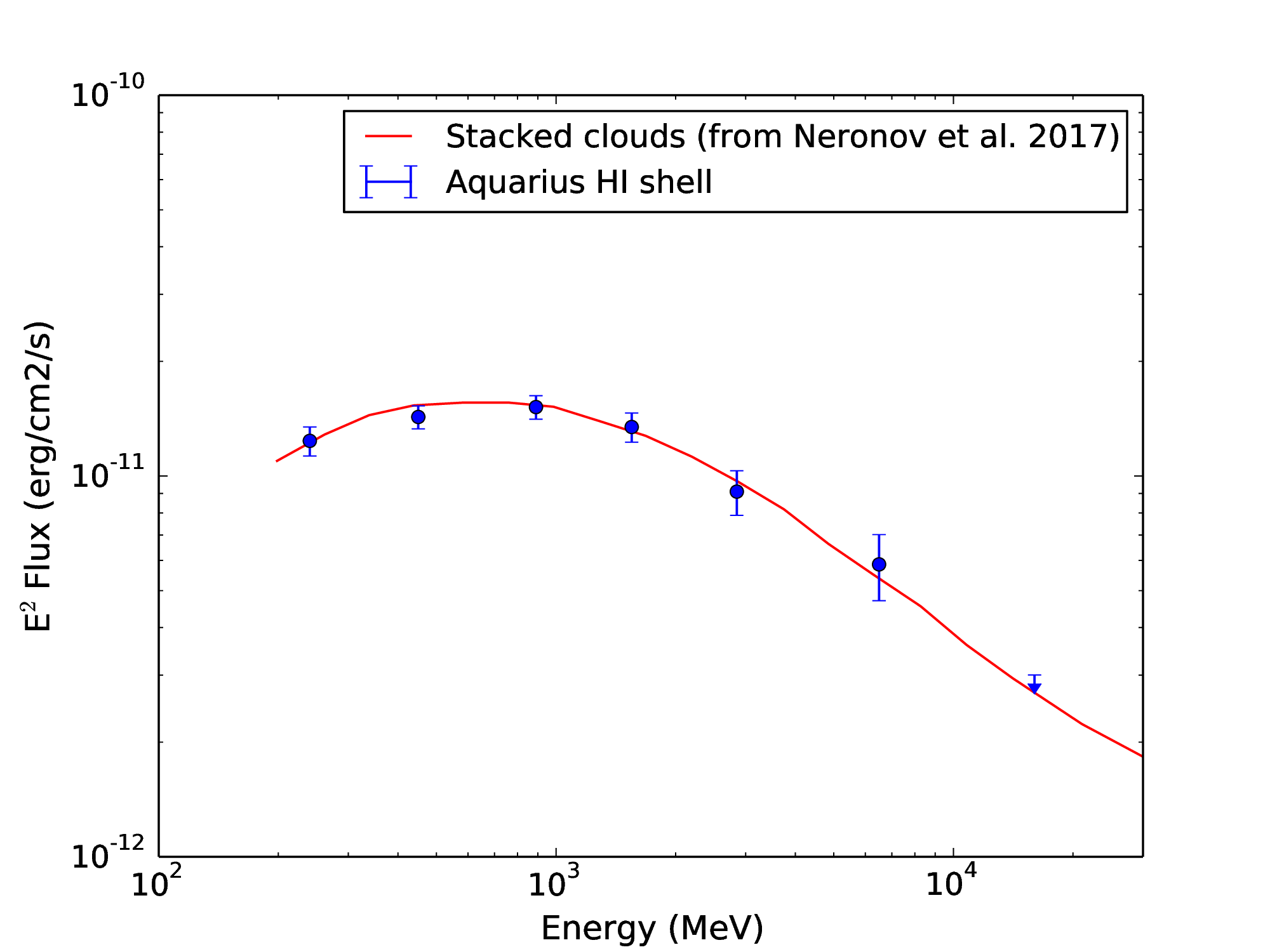}
\caption{The SED of the Aquarius shell as compared with the spectral
model describing the stacked spectrum of the Gould Belt MCs. The
latter model is scaled down by a factor of 120.} \label{F3}
\end{figure}

We computed the spectral energy distribution (SED) for the Aqr
shell. The normalisation in each energy bin for the Aqr shell
component was treated as a free parameter and was obtained as the
result of a likelihood analysis. The computed SED is shown in
Fig.\ref{F3} along with the model derived from the stacked analysis
of the Gould Belt MCs. To compute the upper limit on the
differential flux in the 7th energy bin, we used the
\texttt{UpperLimits} python module. The computed SED agrees well
with the model obtained from the stacked analysis and adopted from
\citet[][]{Neronov17}. We also checked if the spectral slope at high
energies at which most of \gray{} emission from the Aqr shell is
produced by CR protons with energies above 18 GeV is compatible with
that obtained from the stacked analysis of Gould Belt MCs. Since a
typical energy of CR protons producing \grays{} above the minimal
observed energy $E_{\gamma,\mathrm{min}}$ is
$E_{\mathrm{p}}\gtrsim10 E_{\gamma,\mathrm{min}}$, to perform this
check we used the single power-law spectrum to model the Aqr shell's
\gray{} emission above 4 GeV. The best-fitting photon power-law index at
high energies found in our analysis is $2.96\pm0.46$\footnote{In
this case, 87\% of \grays{} with energies above 4 GeV are produced
by CR protons with energies above the break as computed with the
\texttt{naima} python package.} and is compatible with that of
$2.92^{+0.07}_{-0.04}$ obtained from the stacked analysis of the
Gould Belt MCs.

To quantify the difference between the spectrum of CRs producing
the \gray{} signal from the Aqr shell described above and that derived
from \gray{} observations of the R CrA MC by \citet[][]{Neronov17},
we computed both the weighted mean ratios, $R_{1-3}$, for the Aqr
shell fluxes to the stacked MC's modelled fluxes,
$R_{\mathrm{Aqr/Stack}, 1-3}$, and for the Aqr shell fluxes to the R
CrA MC modelled fluxes, $R_{\mathrm{Aqr/RCrA}, 1-3}$. The computed
ratios are $R_{\mathrm{Aqr/Stack}, 1-3}=(7.98\pm0.34)\times10^{-3}$
and $R_{\mathrm{Aqr/RCrA}, 1-3}=0.639\pm0.027$, respectively.
These magnitudes also reflect the fact that the \gray{} flux of
the Aqr shell is
lower than that of the R CrA MC, while the stacked analysis of the
Gould Belt MCs provides us with significantly higher statistics. We
compared the $R_{\mathrm{Aqr/Stack}, 1-3}$ ratio with the ratios of
the Aqr shell fluxes to the stacked MC's modelled fluxes,
$(8.50\pm1.14)\times10^{-3}$ and $(8.99\pm1.77)\times10^{-3}$,
obtained from the higher energy bins centered at 2.85 GeV and 6.5
GeV, respectively. On the basis of the estimators,
we found that the CR spectrum producing \grays{}
from the Aqr shell is compatible within errorbars with the spectrum
of CR populations responsible for most of \grays{} from the Gould Belt
MCs. We also compared the $R_{\mathrm{Aqr/RCrA}, 1-3}$ ratio with
the ratios of the Aqr shell to the modelled R CrA MC fluxes,
$0.74\pm0.10$ and $1.01\pm0.20$, obtained from the higher energy
bins centered at 2.85 GeV and 6.5 GeV, respectively. We found that
the former value is compatible with that obtained at lower energies,
whereas the latter computed ratio exceeds that obtained at lower
energies with about 2$\sigma$ confidence.

We concluded that no variation in the spectrum of CRs above energy of 18
GeV compared with the stacked spectrum of the MCs was derived from
the analysis of the Aqr shell. Since the Aqr shell is located near
the $\rho$ Oph and R CrA MCs and is projected onto the region of a low background,
it provides us with a probe of the CR spectrum in the region containing
these three objects which is less affected by background uncertainties.

\subsection{R CrA molecular cloud}

As mentioned above, the previously derived CR spectrum of the R CrA
MC deviates from the stacked MC's spectrum and this fact was
interpreted as marginal evidence for a departure of the CR spectrum
local to this MC from the CR background spectrum. Two possible
problems in derivation of the CR spectrum of this MC are that (a)
its \gray{} flux is faint in comparison with the Gould Belt MCs'
fluxes and that (b) its position is projected onto the southern
Fermi bubble. Both these problems can be alleviated by using a more
advanced background model (for details on background models, see
Sect. 3.2). We consider two models of background \gray{} emission:
models A and B. Both these models take the gradient of the diffuse
Galactic \gray{} emission into account, but contain different
contributions from the southern Fermi bubble to the R CrA MC \gray{}
signal.

\begin{figure}
\centering
\includegraphics[angle=0,width=9.5cm]{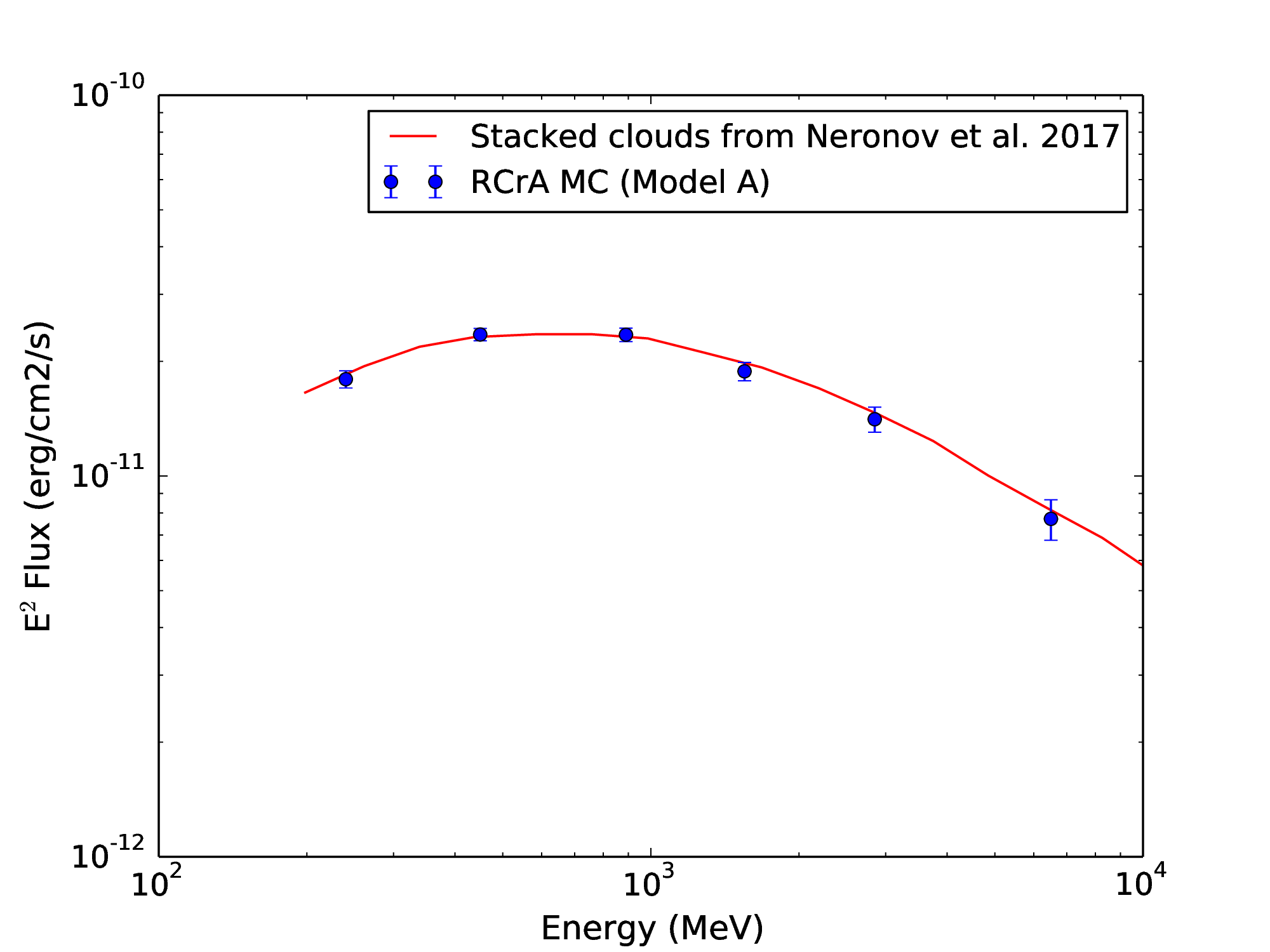}
\caption{The SED of the R CrA MC derived for the background model A
as compared with the spectral model describing the stacked spectrum
of the Gould Belt MCs. The latter model is scaled down by a factor
of 80.} \label{F4}
\end{figure}

We performed a binned likelihood analysis to obtain statistical significances
and differential fluxes of the R CrA MC for the model A.
The evaluated TS values are 507, 1027, 862, 462, 248, and 91 in the six energy
bins. We found that the shape of
the derived \gray{} SED is compatible with that of the \gray{} SED
previously obtained from the stacked analysis of the high latitude
Gould Belt MCs. It suggests that faintness of the R CrA MC in
\grays{} was an obstacle in derivation of its spectrum and supports
that production of \gray{} emission from the R CrA MC involves
hadrons from the CR background. The derived SED is shown in Fig.
\ref{F4}.

We also checked and found that the unassociated source of 3FGL J1902.3-3702c
which is on top of the R CrA MC is detected with only 4.2$\sigma$ in
the analysis based on the model A (and with 5.5$\sigma$ in the analysis
based on the model B) and only slightly affects the results of
the spectral analysis. Compatibility of the results obtained from
the \textit{Fermi}-LAT observations of the Aqr shell and the R CrA MC
(derived in the framework of the model A for the R CrA MC) shows that
the hypothesis of uniformity of the CR spectral shapes does not contradict
these results.

\begin{figure}
\centering
\includegraphics[angle=0,width=9.5cm]{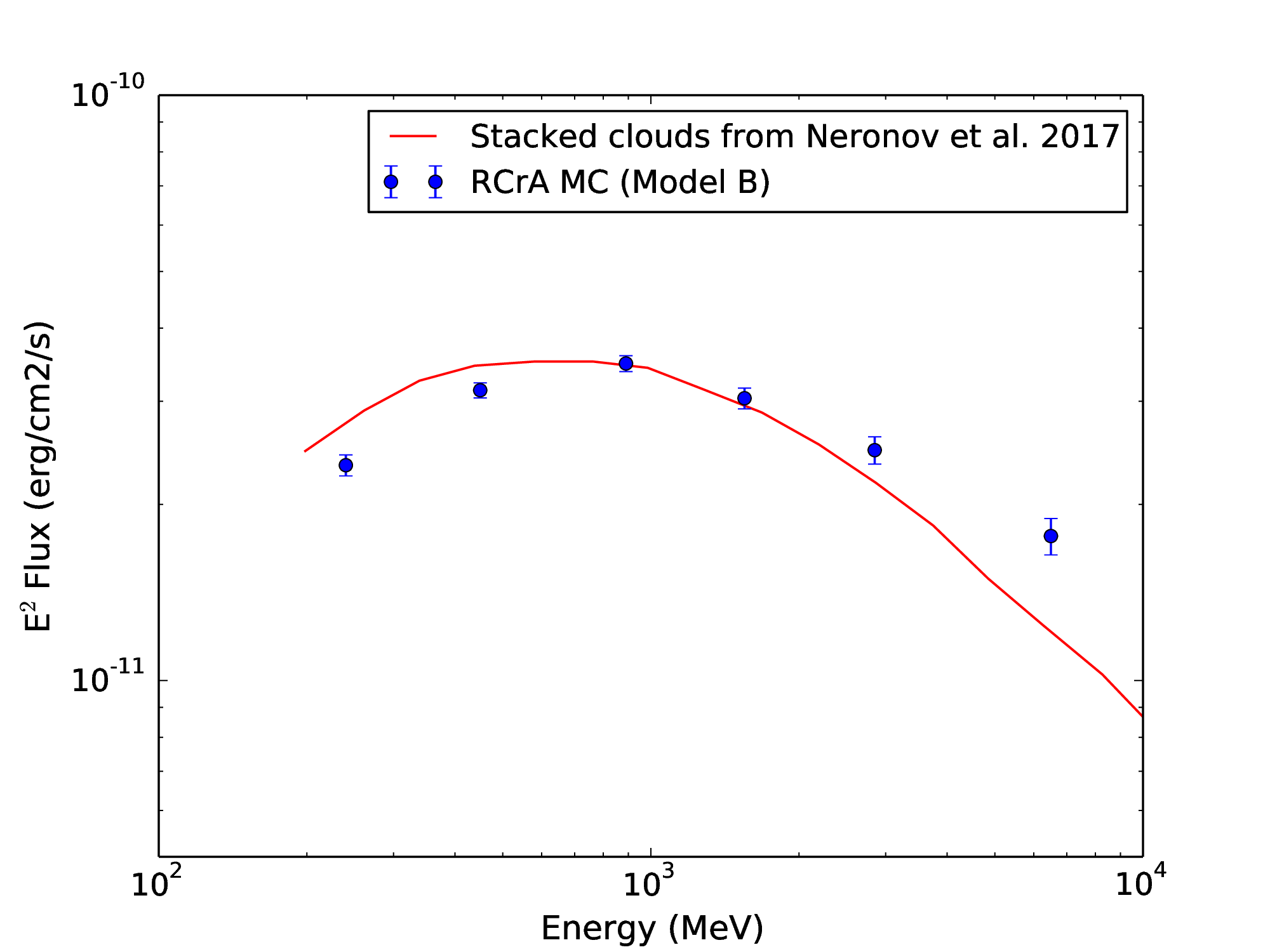}
\caption{The SED of the R CrA MC derived for the background model B
and contaminated by \gray{} emission from the southern Fermi bubble
is shown along with the spectral model describing the stacked
spectrum of the Gould Belt MCs. The latter model is scaled down by a
factor of 53.} \label{F5}
\end{figure}

\subsubsection{A uniform CR spectrum in the local ISM as a tool for
foreground \gray{} subtraction}

The Galaxy is transparent to \grays{} with the exception of rare
events of solar occultations of several \gray{} sources \citep[][]{occ14}.
At high Galactic latitudes, \grays{} produced in the local ISM
involving the hadronic interactions of CRs are the most dominant
foreground component.
In an investigation which requires disentanglement of a
signal of astrophysical origin from a foreground diffuse \gray{}
emitting structure \citep[e.g., for a search for \gray{} emission
from the Virgo cluster which is projected onto the Galactic cirrus,
see][]{Prokhorov14, Bianchi17}, one would be aided greatly
if the CR spectrum was spatially uniform in the local ISM.

We showed the \gray{} spectrum of the R CrA MC as derived from the
model B in Fig.~\ref{F5}. This spectrum strongly deviates from the
stacked MC's modelled spectrum and is harder than that obtained by
means of the model A. The hardness of the spectrum of the R CrA MC
obtained from the model B is most likely explained by the
significant contribution of \grays{} from the southern Fermi bubble
to the R CrA template produced for the model B. It also demonstrates
systematic uncertainties caused by the selection of a background
extraction region for the R CrA MC analysis. Despite these
complexities in the analysis, the fact that the \gray{} spectrum
derived from the model A is compatible with that of the stacked MC's
modelled spectrum likely implies both that (a) the CR spectrum of the
R CrA MC is dominated by the CR background and that (b) the southern
Fermi bubble's cocoon contributes to the total signal from the
square region of side 6$^{\circ}$ enclosing this MC. As note
by \citet[][]{Neronov17}, the exclusion of the R CrA MC from their
sample significantly reduces the amount of evidence for a spatial
variation of the CR spectra in the local ISM. Since we showed that
the derived spectrum of the R CrA MC depends on assumptions about a
$\gamma$-ray background, the exclusion of this MC from their sample
increases reliability of the conclusion.

\subsection{$\rho$ Ophiuchi MC}

\begin{figure}
\centering
\includegraphics[angle=0,width=9.5cm]{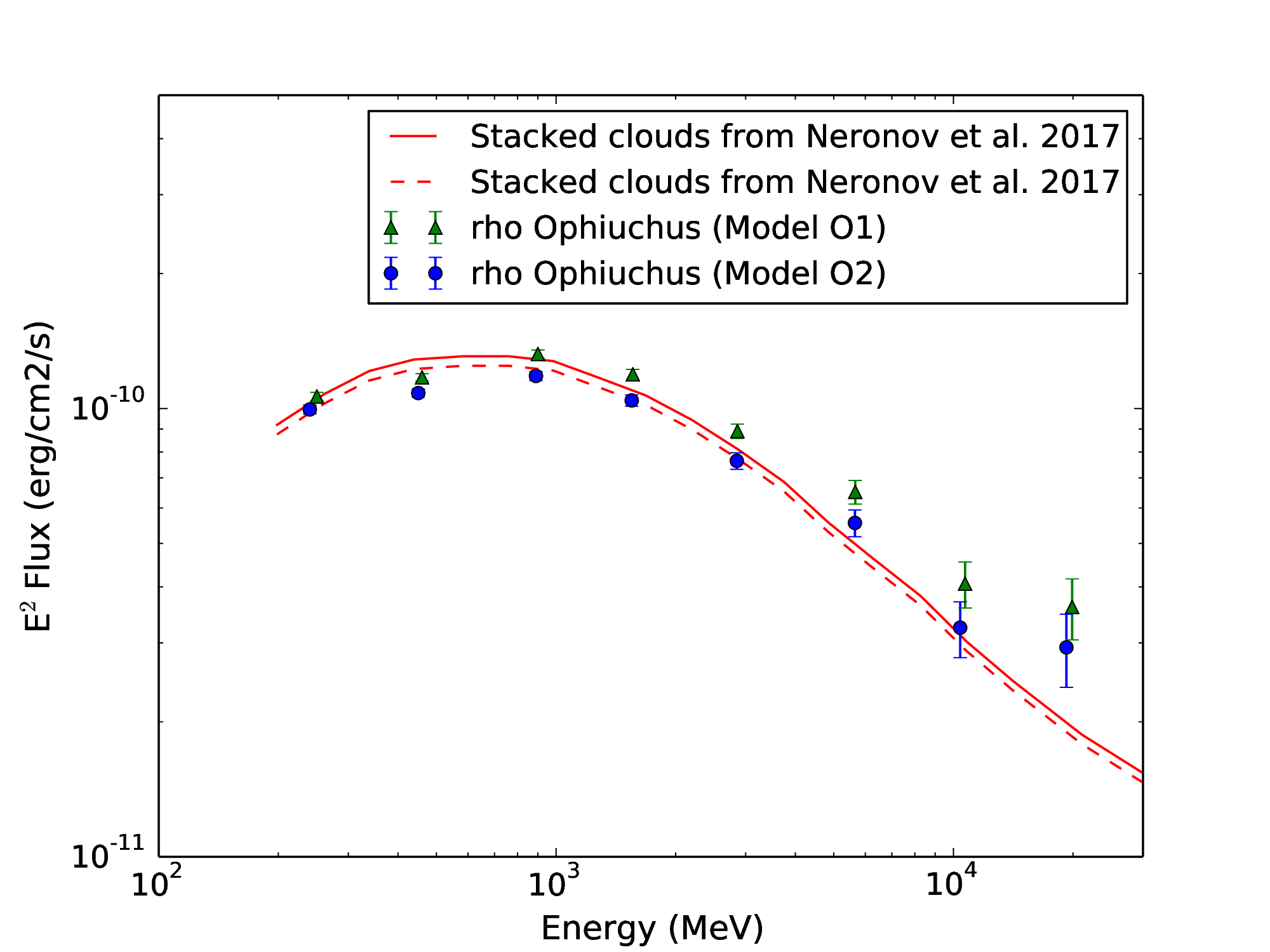}
\caption{The SED of the $\rho$ Oph MC derived for the background models O1
and O2. For the sake of illustration the central positions of the bands
are slightly shifted for the model O2 with respect to those for the model O1.}
\label{FigOph2}
\end{figure}

We used two models of background \gray{} emission, models O1 and O2,
to compute the SED of the $\rho$ Oph MC using a binned likelihood
analysis. Both the models O1 and O2 take the gradient of the diffuse
Galactic \gray{} emission into account. However the latter model
(i.e. O2) also minimises the contribution from the northern Fermi
bubble to the \gray{} signal from the $\rho$ Oph MC. In addition to
the low energy bins, we introduced the high energy bins centered at
energies of 5.7, 10.4, and 19.3 GeV instead of the single bin
centered at 6.5 GeV. We treated the normalisation in each energy bin
for the $\rho$ Oph MC component as a free parameter. We showed
the computed SED in Fig.\ref{FigOph2} for the models O1 and O2. We
also showed the curves taken from the previous stacking analysis of
the MCs and rescaled them for ease of comparison. The SED computed
for the model O1 seems to be harder at high energies than that which
computed for the model O2. To quantify the spectral hardness, we
used the estimators introduced above.

We computed the weighted mean ratios, $R_{1-3}$, for the $\rho$ Oph
MC fluxes to the stacked MC's modelled fluxes, $R_{\mathrm{\rho
Oph/Stack}, 1-3}$, in the models O1 and O2. The computed ratios are
$R_{\mathrm{\rho OphMO1/Stack}, 1-3}=0.068\pm0.001$
and $R_{\mathrm{\rho OphMO2/Stack}, 1-3}=0.064\pm0.001$, respectively.
We found that the $R_{\mathrm{\rho OphMO1/Stack}, 1-3}$ ratio obtained from
the model O1 is in tension with the ratios of $R_{\mathrm{\rho
OphMO1/Stack}, 7}=0.088\pm0.010$ and $R_{\mathrm{\rho OphMO1/Stack},
8}=0.120\pm0.019$ obtained from the higher energy bins centered at
10.4 and 19.2 GeV in the framework of the same model. The tension
increases to 2.7$\sigma$ statistical level in the energy bin
centered at 19.2 GeV. The advanced model, O2, reduces the deviation
from the stacked MC's spectrum and the $R_{\mathrm{\rho
OphMO2/Stack}, 1-3}$ is compatible
with the ratios of $R_{\mathrm{\rho OphMO2/Stack}, 7}=0.071\pm0.010$
and $R_{\mathrm{\rho OphMO2/Stack}, 8}=0.098\pm0.018$ obtained from
the higher energy bins on the 2 $\sigma$ statistical level.

We concluded that the model O2 which minimises the contribution from
the northern Fermi bubble to the $\rho$ Oph MC \gray{} signal allows
us to describe observations of the $\rho$ Oph MC collected by
\fermilat{} during the first 4 years of the mission in the framework
of the standard CR background scenario for \gray{} production.
It is worth to note that only the $\rho$ Oph and R CrA MCs
from the sample of MCs from \citet[][]{Neronov17} both are projected
onto the Fermi bubbles and show spectral deviations compared with
the stacked spectrum of the MCs, while the other MCs from that
sample are projected away from the Fermi bubbles and do not show any
spectral deviations. We suggest that the most probable explanation
of the spectral deviations related to the $\rho$ Oph and R CrA MCs
is a difficulty in background modelling. Given that the Aqr shell is
located near the $\rho$ Oph and R CrA MCs and is projected away
the Fermi bubbles, the compatibility of the $\gamma$-ray spectrum of
the Aqr shell and the stacked spectrum of the Gould Belt MCs
provides evidence to support the hypothesis of uniformity of the
shapes of CR spectra in the local Galaxy environment.

\section{Discussion and outlook}

If one SN event injects $E_\mathrm{inj}\sim10^{50}$ erg in CRs, it
might produce a significant increase in the overall CR flux above 18
GeV in the ISM gas structure located at $R=150$ pc from the SN site.
To show this, let us assume that the intensity of cosmic
rays to be the one measured near the Earth,
$J(E)=1.8\times\left(E/\mathrm{GeV}\right)^{-2.7}$
cm$^{-2}$s$^{-1}$sr$^{-1}$GeV$^{-1}$ \citep[][]{Gaisser90} and
calculate the integral of the intensity above $E_{\mathrm{brk}}=18$
GeV. One finds that the CR energy density above $E_{\mathrm{brk}}$,
$\epsilon=(4\pi/c)\times\int^{\infty}_{E_{\mathrm{brk}}} J(E) E
\mathrm{d}E$, is $\epsilon=2.3\times10^{-13}$ erg cm$^{-3}$. One SN
event is sufficient to fill the volume of a sphere of radius,
$R=\left(3 E_{\mathrm{inj}}/(4\pi \epsilon)\right)^{1/3}$, which is
about 150 pc, with CR protons of energy density similar to the
locally measured one. These CR spectral distortions can be possibly
detectable in \grays{} with \textit{Fermi}-LAT.
This calculated distance determines the radius of influence
of one SN event. It is expected that the ISM structures at a mutual
distance of less than the radius of influence have a similar imprint
on their high-energy CR spectra left by one SN event.
Therefore, indirect CR measurements by means of $\gamma$-ray
observations in such ISM structures, including those which we
analysed in the paper, are of interest to search for a signature of
discrete source CR injection.

In the case of anisotropic propagation of CR protons, the radius of
influence can depend on the direction because CRs spread faster
along the Galactic magnetic field lines. Owing to this
effect, the CR spectra in the nearby ISM gas structures could
potentially be different from each other. Thus, the direction of
magnetic field lines in the space between three nearby ISM
structures would be measured if their CR spectra were different.
This is because any three non-collinear points in space
determine a unique plane. A normal vector to the plane determines
the direction perpendicular to a line between any two of given three
points.

Radio band and \gray{} observations of nearby ISM gas structures are
a unique tool to look for the effect of discreteness of CR injection
events in space and time. Accumulation of \grays{} and mapping of
local ISM structures with radio observations will open the
possibility for a further analysis including disentanglement of the
contributions from the interacting HI shell and MC in each of these
analysed structures as well as in other local regions of the Galaxy
and possibly a determination of the direction of a
magnetic field in the region of three ISM structures localised
within the radius of influence.

\section{Conclusions}

The p-p collisions between CR hadrons and the ISM gas followed by neutral pion
decay result in production of \grays{} measured with telescopes,
e.g. \textit{Fermi}-LAT. The ISM gas structures containing
atomic and/or molecular hydrogen provide a target for such collisions.
The spatial distribution of atoms or molecules of hydrogen is available from
observations of HI 21-cm or CO 2.6-mm lines with radio telescopes, respectively.
Measurements of \gray{} spectral shapes produced in the ISM gas accumulations
can make a study of injection sites of CR protons possible.

The first 3 years of \textit{Fermi}-LAT \gray{} observations of high
latitude Gould Belt MCs led to the conclusion that the spectra of
individual MCs are consistent with each other \cite[][]{Neronov12},
while 8 years of \textit{Fermi}-LAT observations of the same MCs led
to the indication of variations in the slope of the CR spectrum
above about 18 GeV \cite[][]{Neronov17}. A transition from the
steady-state continuous injection to the regime of discrete source
injection was proposed as an explanation for the slope variation in
CR spectra of these MCs. They found that the softest and hardest
spectral slopes at high energies are in the R CrA MC and in the
$\rho$ Oph MC, respectively. Comparing the mutual distances between the MCs
from their sample, we found that the distance between the R CrA and
$\rho$ Oph MCs is one of the shortest (after that between the Orion A
and Orion B clouds) and is only 100 pc. Given the close mutual
distance of these two clouds with distinct CR spectra, we searched
for another ISM structure located in the vicinity of these clouds to
test similarity of CR spectra in this region. We found that the HI
shell in Aquarius separated from the R CrA cloud by 110 pc
provides us with an alternative probe of the CR spectrum
in the region surrounding the R CrA and $\rho$ Oph MCs and the Aqr shell.
We derived the SED of \gray{} emission from the Aquarius shell and
tested the shape of the CR proton spectrum. The tests showed that
the  CR slope obtained in the Aqr shell is compatible with that
obtained from the stacked spectrum of Gould Belt MCs. We also
re-analysed the \gray{} emission from the R CrA and $\rho$ Oph MCs.
We found that a more detailed background model including both the
Galactic \gray{} emission gradient and the \gray{} emission from the
Fermi bubbles allow us to reconcile the derived \gray{} spectrum of
these two nearby MCs with that obtained from the stacked analysis of
MCs. Taking into account that only these two MCs from the sample
taken from \citet[][]{Neronov17} are projected onto the Fermi bubbles,
the most probable explanation of the previously claimed spectral deviations
in these two MCs is a difficulty in background modelling. All this provides
evidence to support the hypothesis of uniformity of the CR spectral shapes
in the local Galaxy environment.

\section{Acknowledgements}
We are grateful to the referee for the constructive
suggestions that helped us to improve the manuscript. This work is
based on the research supported by the South African Research Chairs
Initiative of the Department of Science and Technology and National
Research Foundation of South Africa (Grant No 77948). D.A.P.
acknowledges support from the DST/NRF SKA post-graduate bursary
initiative.

\bibliography{refs2}

\end{document}